\documentstyle{mn}

\newcommand{\ale}{\ \raisebox{-.3ex}{$\stackrel{<}{\scriptstyle \sim}$}\ }
\newcommand{\age}{\ \raisebox{-.3ex}{$\stackrel{>}{\scriptstyle \sim}$}\ }
\newcommand{\omb}{\Omega_{\rm orb}}

\newcommand{\vsc} {$d\,\Omega_{\rm orb}$}
\newcommand{\lsc} {$d$}
\newcommand{\tsc} {$\Omega_{\rm orb}^{-1}$}

\input psfig
\title[Precession of eccentric discs]{The precession of eccentric
discs in close binaries}
\author[J.R. Murray]{J.R. Murray\thanks{email: jmu@star.le.ac.uk}\\
Department of Physics \& Astronomy, University of Leicester,
University Road, Leicester LE1 7RH, UK}
\begin{document}

\maketitle

\begin{abstract}
We consider the precession rates of eccentric discs in close binaries,
and compare theoretical predictions with the results of numerical disc
simulations and with observed superhump periods. A simple dynamical
model for precession is found to be inadequate. For mass ratios $\ale
1/4$ a linear dynamical model 
does provide an upper limit for disc precession rates. 
Theory suggests that
pressure forces have a significant retrograde impact upon the
precession rate (Lubow 1992). We find that the disc precession rates
for three systems with accurately known mass ratios are significantly
slower than predicted by the dynamical theory, and we attribute the
difference to pressure forces. By assuming that  pressure
forces of similar magnitude occur in all superhumping systems, we obtain an improved fit to
superhump observations. 
\end{abstract}


\section{Introduction}
Superhumps are now commonly found in short period cataclysmic
variables (Patterson 1998) and  X-ray binaries (O'Donoghue \& Charles 1996).
The most likely explanation is that these periodic luminosity
variations   are caused by the 
tidal stresses on an eccentric, precessing accretion disc (Whitehurst
1988; Hirose \& Osaki 1990; Lubow 1991). In this model, 
the superhump period, $P_{\rm sh}$, 
equates to the period of the disc's precession, $P_{\rm d}$, 
as measured in the {\em binary} frame. Although reliable measurements
of $P_{\rm sh}$ have been made for at least 53 separate systems
(Patterson 1998), the observations have yet to be compared properly
with the model's theoretical predictions. 

In this paper we consider the factors determining the precession
rate of an eccentric disc, and compare our theoretical understanding, 
simulation results and the observations.

\section{disc precession periods}
It is now well established that in binaries with mass ratios 
$q \ale 1/4$ it is possible for eccentricity to be excited in the accretion
disc at the $3:1$ eccentric inner Lindblad resonance (Lubow 1991). Here we
define $q=m_{\rm 2}/m_{\rm 1}$ to be the mass of the donor star
divided by the mass of the accreting star. Under certain
circumstances, eccentricity may be excited in systems with mass ratios
as large as $1/3$ (Murray, Warner \& Wickramasinghe 1999).

How an eccentric disc precesses coherently has been a source of confusion.
For a single particle, the rate at which an elliptical orbit precesses
\begin{equation}
\omega_{\rm dyn}=\Omega-\kappa.
\label{eq:dyn}
\end{equation}
Now, for a given gravitational potential, a particle's angular
frequency, $\Omega$, and   radial or
epicyclic frequency, $\kappa$, are both functions of $r$. Hence the precession
rate, $\omega$, also depends upon the mean radius of the orbit.  
Yet neither the observations (Patterson 1998) 
nor the simulations (Murray 1998) yield any indication of
differential precession. So how does a disc, that extends over a 
range of radii, organize itself to precess with a unique frequency?

The answer is that we are not dealing with a collection of {\em
isolated} test particles, but with a gaseous disc. 
The eccentricity is excited at the Lindblad resonance, and then
propagates inwards through the disc as a wave,
getting wrapped into a spiral by the differential rotation of
the gas (Lubow 1992). 
If the spiral is wound up on a length scale  much smaller than the radius
(the tight winding limit), then  
$\omega$, the rate of azimuthal advance of the eccentricity, 
is governed  by the dispersion relation
\begin{equation}
(\Omega-\omega)^2= \kappa^2 + k^2c^2,
\end{equation}
where $c$ is the sound speed of the gas, and $k$
is the radial wavenumber of the spiral (see e.g. Binney \& Tremaine 1987).
 Now, as $\omega$ is much less than $\Omega$ and $\kappa$ we have
\begin{eqnarray}
\omega &\simeq& \Omega - \kappa - k^2c^2/(2\Omega)\\
&=&\omega_{\rm dyn} - k^2c^2/(2\Omega).
\label{eq:totprec}
\end{eqnarray}
$\omega$ is determined for the wave in the region in which it is launched.
As the wave propagates inwards, the frequency remains constant but 
the wavelength changes in
response to the changing environment through which it moves.

The above arguments are based upon the analysis of Lubow (1992). He
 also identified a third factor that contributed
to the precession when the magnitude of the eccentricity was changing
 secularly. In this paper we are interested in the long term mean
 value for $\omega$ and so this extra term need not be considered further.

Equation~\ref{eq:totprec} tells us that a gaseous disc will precess
{\em more slowly} than a ballistic particle at the resonance radius.
 Lubow (1992) estimated that pressure effects could reduce
$\omega$  by approximately one percent of $\Omega$.  
Now $\omega_{\rm dyn}$ is itself of the order of a few per cent of
$\Omega$, so the reduction is significant. (Simpson \& Wood (1998)
misinterpreted Lubow's work and ignored the pressure term as being a
few percent of the dynamical precession).

Hirose \& Osaki (1993) obtained estimates for the superhump period by
solving the eigenvalue problem of a linear one-armed mode in an
inviscid disc. Their arguments correspond with those given above
and their results agree very well with those of Lubow (1992).
They showed that an increase in disc temperature allowed the eccentric
mode to propagate further into the disc, with the precession rate
being reduced as a result.
Lubow (1992) also completed
several two dimensional hydrodynamic 
disc simulations, in which he isolated the various
contributing factors to the precession. His numerical results clearly showed
pressure forces to be important. In fact, for those particular
simulations, the pressure contributions reduced $\omega$ to half the
dynamical value. Till now, observational data has only been compared
with dynamical estimates for the precession that we know to be inadequate.

\section {The theory and simulations compared}
The numerical simulations of Whitehurst (1988) lead directly to the
eccentric disc model for superhumps.  
In this section we compare subsequent numerical results (Hirose \& Osaki
1990; Kunze, Speith \& Riffert 1997; Murray 1998; Simpson \& Wood 1998;
Murray et al. 1999)
with the dynamical equation for precession (equation~\ref{eq:dyn}). 
We do not attempt a comparison with the hydrodynamic equation as
the simulations of Hirose \& Osaki  were completed
using the sticky particle method due to Lin \& Pringle (1976) which
does not account for pressure forces. 

\begin{figure}
\psfig{file=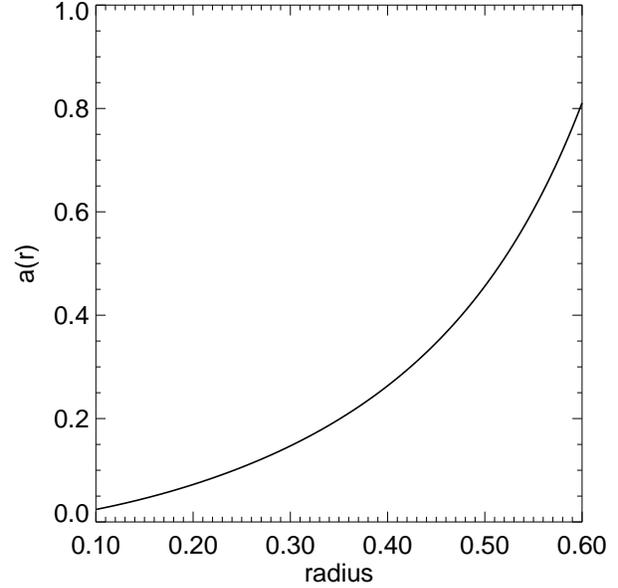,width=8cm}
\caption{$a(r)$ is the term encapsulating the radial dependence of the
dynamical precession rate. Calculated using the 
hyper-geometric series (Brouwer \& Clemence, 1961).}
\label{b0}
\end{figure}

For approximately circular orbits, the rate of dynamical precession
\begin{equation}
\omega_{\rm dyn}=a(r)\,\frac{q}{\sqrt{1+q}}\,\omb,
\label{eq:dyndetail}
\end{equation}
where
\begin{equation}
a(r)=\frac{1}{4\,r^{1/2}}\,\frac{d}{dr}(r^2\,\frac{db^{(0)}_{1/2}}{dr}).
\end{equation}
$b^{(j)}_s$ is the standard notation for a Laplace coefficient from celestial
mechanics.  
We have made use of the hyper-geometric series expression for 
$b^{(j)}_s$ found in Brouwer \& Clemence (1961) (equation 42, Chapter 15)
to evaluate $a(r)$ as a function of radius (see
figure~\ref{b0}). Clearly, 
the precession rate of a ballistic particle is an increasing
function of radius. But as mentioned above, differential precession is
not observed. Thus in previous applications of the dynamical theory
to discs, it was assumed that the entire disc precessed as if it were
a single ballistic particle at the resonance radius, 
$r_{\rm res}\simeq 0.477\,$\lsc, (see e.g. Hirose \& Osaki).
Unfortunately, most of these comparisons are of little value because they
either failed to correctly evaluate $a(r)$ (e.g. Patterson 1998) or
they made use of an incorrect equation for the precession that first appeared
in Whitehurst \& King (1991).  

\begin{figure}
\psfig{file=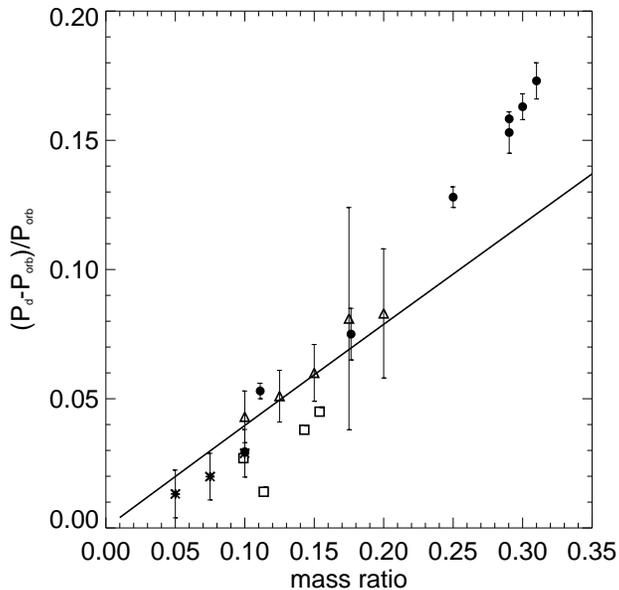,width=8cm}
\caption{Collation of superhump period excesses from;  
Murray (1998), Murray et al. (1999) and this paper
(filled circles), Hirose \& Osaki (triangles),
Simpson \& Wood  (asterisks), and Kunze et al.  (hollow squares).
The solid curve is the dynamical
contribution to the period excess at the resonance radius
(as calculated using equation~\ref{eq:dyndetail})
from this paper, Simpson \& Wood, and Kunze et al. lie very close to
one another on this plot.}
\label{shall}
\end{figure}

Observational data is usually presented in terms of the superhump period
excess
\begin{equation}
\epsilon=\frac{P_{\rm sh}-P_{\rm orb}}{P_{\rm orb}}.
\end{equation}
In terms of the disc precession rate $\omega$, 
\begin{equation}
\epsilon=\frac{\omega}{\omb-\omega}.
\end{equation}
So in figure~\ref{shall} we have plotted as a function of mass ratio the 
superhump period excesses obtained from the disc simulations of
several different authors. The solid curve shows the 
superhump period excess estimated by equation~\ref{eq:dyndetail} 
with $r=r_{\rm res}$.

All authors used the smooth particle hydrodynamics (SPH) technique, except
 Hirose \& Osaki  who used the sticky particle technique.
The
pressure contribution  to the disc precession
is thus  absent from their simulations. The various implementations of
SPH are described in Flebbe et al. (1994), Simpson \& Wood (1998),
and Murray (1996).

The calculations completed for Murray (1998), 
and Murray et al. (1999)
were of very cool isothermal discs. In units of the binary separation,
$d$, and the orbital angular frequency, \tsc, we set the sound speed
$c=0.02\,$\vsc. 
The pressure contribution is proportional to the square of the sound speed
and so was very small. We also used a very large value for the shear
viscosity. This  reduced the viscous time scale to a value
that enabled us to follow the evolution of discs to 
steady state in a reasonable amount of
computing time. However with a larger shear viscosity it was easier
for material to penetrate the resonance. Significantly eccentric discs
and very strong superhump signals
were obtained as a result. Furthermore,  the large shear viscosity
inhibited the inward
propagation of the eccentricity and further reduced the effectiveness of the
pressure term in equation~\ref{eq:totprec}.

Without the benefit of any significant pressure contribution to the 
disc precession, 
the points from Murray (1998) and Hirose \& Osaki lie very close to
the dynamical precession curve. Some  are marginally
above the curve. As mentioned above, the large shear
viscosities used allowed the discs to effectively penetrate the
resonance and extend to somewhat larger radii.  
On the other hand the four points from
Murray et al. (1999)  
are all well above the curve, and cannot be
explained in terms of large discs. 
For these mass ratios the resonance lies beyond the tidal
truncation radius. That is, for $q \age 1/4$ 
simply periodic orbits begin intersecting one
another inside $r_{\rm res}$. The orbits in this region will no longer
be approximately circular, in contradiction of the assumption underlying the
analytical expressions in Hirose \& Osaki, and Lubow (1992). We
conclude therefore that, gas pressure considerations aside, expressions such as
equation~\ref{eq:dyndetail} will only be valid  
for mass ratios less than approximately 1/4. 

The discs of Simpson \& Wood, and of Kunze et al.
precessed significantly more slowly than expected
from purely dynamical considerations (these points lie below the
dynamical curve in figure~\ref{shall}).
Kunze et al. assumed each element of the disc radiated as a blackbody
and set the sound speed according to the blackbody
temperature. 
Simpson \& Wood  used a polytropic equation of state with 
index $\gamma=1.01$ and integrated the internal energy equation for
each particle.  Although a comparison is not straightforward,
these discs 
would have been somewhat hotter than the ones constructed for this
paper. As a consequence,
the retrograde contribution to the precession due to pressure
forces would have been larger, and the values for $P_{\rm d}$ obtained by
Simpson \& Wood  and Kunze et al. are closer to observed $P_{\rm sh}$.  

We are currently running a set of
simulations with the shear viscosity ten times smaller than in Murray
(1998), but with other parameters unchanged. In particular, the sound
speed $c=0.02\,$\vsc. This gives an effective value of the Shakura-Sunyaev
parameter $\alpha\simeq 0.16$ at the resonance (as opposed to $\simeq
1.6$ in the previous calculations).
At present we have
results only for $q=0.10$ (the simulations now take much longer to
complete). This particular calculation ran for 100 orbital periods.
As is to be expected, the superhump amplitude was reduced.
At the conclusion of the calculation the 
superhump period (averaging over 25 superhump cycles) $P_{\rm
sh}=(1.0295\pm 0.0005)\,P_{\rm orb}$. 
This period is significantly below the dynamical precession
curve even though the disc is very cool, and it corresponds very well
with the results of Simpson \& Wood, and Kunze et al.
Our result has been included in figure~\ref{shall} but we
emphasize that at the conclusion of the simulation, the disc mass was
still very slowly increasing, and the superhump period had not
completely stabilised.

In the above discussion we did not refer to the calculations of
Whitehurst (1994) simply because he used an initial  mass transfer
burst to set up his discs, and the superhump periods were not
obviously steady state values. The simulations described in section
4.2 illustrate the time scale upon which resonant discs adjust to
changes in the mass transfer rate. Before we leave this subject, a
word of warning.
 We discovered in one particular trial
simulation that once mass return from the outer edge of the disc to
the secondary star occurred, the superhump period excess was reduced
by approximately $5\,\%$. The explanation is simple. Material at the
outer edge of the disc precesses most rapidly. Remove that material
and you slow the disc precession. This effect 
can hinder comparison with theory and other simulations.  Such mass
return occurred in the simulations of Hirose \& Osaki  and of
Simpson \& Wood. 

We conclude that for mass ratios less than $1/4$,
equation~\ref{eq:dyndetail} provides a useful {\em upper limit} for
the steady state precession rate of a gaseous disc. However, for
systems with $q \age 1/4$, the intersection of simply periodic orbits
near the resonance radius renders equation~\ref{eq:dyndetail} invalid.

The next step is to compare  simulation results with eigenmode
calculations such as those performed by Hirose \& Osaki (1993). They
tabulated disc precession rates as a function of the sound speed at
the resonance for systems with mass ratios $q=0.15$ and $0.05$.
Both the eigenmode calculations and the hydrodynamic simulations show
similar magnitude pressure contributions to the precession.
However, as yet the 
assumptions made in making the two sets of calculations (e.g. equation
of state) differ sufficiently as to prevent a more detailed comparison.

\section{Theory and observation compared}
Patterson (1998)  tabulated the superhump period excesses for 53
systems.
In this section we will compare this data with both the dynamical and
hydrodynamical equations for precession. In order to do this we
require an independent means of estimating a system's mass
ratio. Unfortunately, 
although
the mass of the secondary star as a function of orbital period is
reasonably well constrained, there is evidence of considerable
variation in the white dwarf masses of otherwise similar systems (see
e.g. Smith \& Dhillon 1998, figure 5). The
resulting uncertainty in $q$ is large enough to interfere with our
comparisons with theory. An early attempt to compare theory and
observation by Molnar \& Kobulnicky (1992), was hampered in exactly
this fashion (see their figure 2).
\begin{figure}
\psfig{file=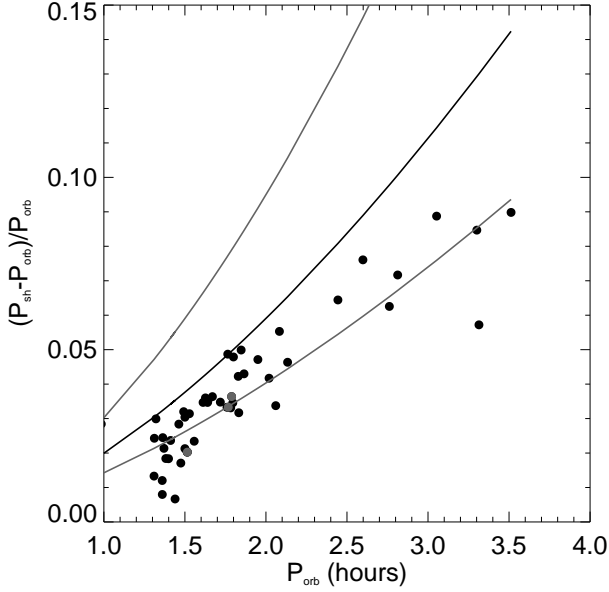,width=8cm}
\caption{Observed superhump period excesses (data from Patterson,
1998) plotted against orbital period. Overlaid are curves for the
dynamical precession as calculated using equation~\ref{eq:dyndetail}, 
and assuming 
 a white dwarf mass $M_{\rm wd}=0.76\pm 0.22 \,M_\odot$, and  the
secondary mass period relationship from Smith \& Dhillon (1999), i.e. 
$M_{\rm sec}/M_\odot=(0.038\pm 0.003)P^{(1.58\pm 0.09)}$. The outlying
curves incorporate the uncertainties in the Smith \& Dhillon values.}
\label{shobs}
\end{figure}


In figure~\ref{shobs} we have plotted the superhump data from Patterson (1998) 
against theoretical predictions for
the dynamical precession (made using equation~\ref{eq:dyndetail} with 
$r=r_{\rm res}$). In order to estimate
the mass ratio of a system of given orbital period we have used the 
observationally derived stellar masses obtained by Smith \& Dhillon
(1998). They found  
the mean white dwarf mass  for all cataclysmic variables to be $0.76
\pm 0.22 \,M_\odot$,  and they estimated that
$M_{\rm sec}/M_\odot=(0.038\pm 0.003)P^{(1.58\pm 0.09)}$. 
The central (darker) curve in figure~\ref{shobs} represents the best
theoretical estimate. The outer two curves show the consequences of
the uncertainty in the Smith \& Dhillon values for the theory.
In fact, Smith \& Dhillon found the mean white dwarf mass for systems
below the period gap to be $0.69\pm 0.13\,M_\odot$, and to be $0.80\pm
0.22\,M_\odot$ for systems above the period gap. However precession rates
recalculated using the adjusted white dwarf masses differed only
marginally from the original estimates. 

Assumptions about stellar masses aside, figure~\ref{shobs} shows that 
the superhump observations 
{\em cannot} be adequately explained in terms of simple dynamical precession. The
retrograde precession due to pressure forces is necessary to bring
closer agreement between theory and observations. Previously published
figures showing closer agreement between dynamical precession
estimates and observations  only did so because
the dynamical precession was calculated incorrectly. For example
Patterson (1998) used the equation
\begin{equation}
\omega_{\rm dyn}/\omb=0.23\,\frac{q}{\sqrt{1+q}}.
\label{patt}
\end{equation}
The coefficient should be approximately $0.4$ 
(see Hirose \& Osaki 1990; Lubow 1992; figure~\ref{b0} above).  In order to explain
superhumps in long period systems with equation~\ref{patt}, 
mass ratios as high as $0.5$ are required. These $q$ values are
clearly incompatible with eccentricity excitation at the $3:1$ inner
Lindblad resonance. However, if we use the more defensible  
equation~\ref{eq:totprec} for the precession, then smaller mass
ratios are obtained.

Our confidence in the comparison between theory and observation is
limited by uncertainty in $q$.  The inadequacy of the dynamical
expression for precession is much more clearly apparent when we
consider the eclipsing systems OY~Car, HT~Cas and Z~Cha. Very accurate
determinations of the mass ratios of these systems are respectively
obtained in Wood et
al. (1989), Horne, Wood \& Stiening (1991), and Wood et al. (1986).
In Table~\ref{tab:wellk} we list for each system the observed $\omega$ and the
 dynamical precession rate. In each case the difference between
the two values is too large to be explained by uncertainties in either
$q$ or equation~\ref{eq:dyndetail}. The difference is simply the
(retrograde) pressure contribution to the precession.   

We can check whether the inferred $\omega_{\rm pr}$ is consistent with
the assumption that the eccentricity is tightly wound.
The pitch angle $i$ of the spiral wave is given by $\cot i=
|kr|$. Thus, substituting for $k$ in equation~\ref{eq:totprec}, 
we find that the pressure
contribution to the precession rate at the $3:1$ resonance
\begin{equation}
\omega_{\rm pr}\simeq-
\frac{2}{3}\,\omb\,(\frac{c}{\omb\,d\,}\frac{1}{\tan{i}})^2. 
\label{eq:lubprec}
\end{equation}
Note that Lubow (1992) had the constant of proportionality inverted in
his equation 21. If we assume the sound speed at the resonance radius
$c = 0.05\,d\,\omb$ then we obtain pitch angles of $17,16$ and
$15^\circ$ for OY~Car, Z~Cha and HT~Cas  respectively. These are
certainly consistent with the tight winding approximation. 

\begin{figure}
\psfig{file=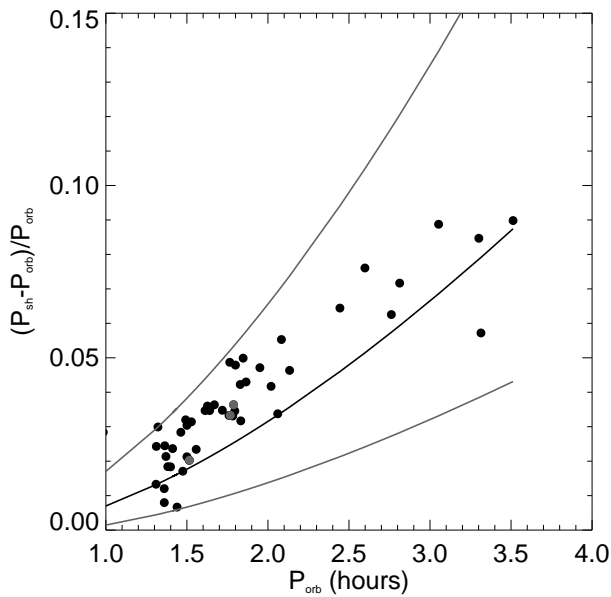,width=8cm}
\caption{Observed superhump period excesses (data from Patterson,
1998) plotted against orbital period. Overlaid are curves for the
hydrodynamic precession as calculated using equation~\ref{eq:totprec},
with $\omega_{\rm pr}= -1.9 {\rm rad}\, {\rm day}^{-1}$. We
assume
 a white dwarf mass $M_{\rm wd}=0.76\pm 0.22\,M_\odot$, and  the
secondary mass period relationship from Smith \& Dhillon (1999), i.e. 
$M_{\rm sec}/M_\odot=(0.038\pm 0.003)P^{(1.58\pm 0.09)}$. The outlying
curves incorporate the uncertainties in the Smith \& Dhillon relation.}
\label{shp}
\end{figure}

We complete this section with a comparison of the observations with
the predictions of the hydrodynamic theory. As it is not clear what
value $k$ should take in equation~\ref{eq:totprec}, we make
the naive assumption that the
pressure contribution to precession in all 53 systems tabulated in
Patterson (1998) will be similar to that of OY~Car, HT~Cas and Z~Cha.
We take a mean value for $\omega_{\rm pr}=-1.9\,{\rm rad}\,{\rm
day}^{-1}$, and plot predicted precession rates against the
observations in figure~\ref{shp}. A much improved fit is
achieved. As with figure~\ref{shobs}, three curves are drawn to show
the influence of uncertainty in our knowledge of white dwarf and
secondary masses. Much of the difference between the curves
is due to the uncertainty in white dwarf mass, with the uppermost
curve corresponding to $M_{\rm wd}=0.54\,M_\odot$. 
The observational data is perhaps
best fit with a curve generated assuming a white dwarf mass $\simeq
0.65\, M_\odot$. 

We recall that equation~\ref{eq:dyndetail} underestimates the
dynamical precession rate for systems with mass ratios $\age 1/4$. 
Thus the theoretical curves should rise more steeply
long-wards of say $P_{\rm orb} = 3$ hr. The implication then is that
long period superhumpers have significantly more massive white dwarfs
than do their shorter period counterparts. As a consequence the mass
ratios of the long period systems will be smaller than previously
estimated.
This is qualitatively consistent with the eccentricity being excited
at the $3:1$
Lindblad resonance, and with our previous result (Murray et al. 1999) 
that the excitation can occur at mass ratios
$\ale 1/3$. In other words, of those systems with say $P_{\rm orb} \simeq
3$ hrs, only those with more massive white dwarfs will exhibit superhumps.
This then is distinct from any effect caused by systematic variation in
white dwarf masses in the cataclysmic variable population as a whole.
\begin{table}
\caption{The observed disc precession rate (in rad ${\rm day}^{-1}$)
for three systems with accurately known mass ratios. The dynamical
precession rate is calculated using equation~\ref{eq:dyndetail}
The pressure contribution is then simply the difference between those
two values.}
\label{tab:wellk}
\begin{tabular}{ccccc}
System & $q$ & $\omega_{\rm obs}$ & $\omega_{\rm dyn}$ & $\omega_{\rm pr}$\\
\hline
OY~Car &$0.102\pm 0.003$ & 1.977 & 3.869 & -1.892\\
Z~Cha &$0.15\pm 0.03$ & 2.961 & 4.712 & -1.751\\
HT~Cas &$0.15\pm 0.03$ & 2.75 & 4.773 & -2.02\\
\hline
\end{tabular}
\end{table}

\section{Conclusions}
We have compared the theoretical predictions for the precession rates
of eccentric discs with simulation results and with observed superhump
periods. Comparison with disc simulations showed that the dynamical
equation provided a useful upper limit for the disc precession  rate
of systems with mass ratios $\ale 1/4$. 

Previous papers showing good agreement between observations and a
simple dynamical model for precession are based on incorrectly calculated
dynamical expressions. The consequence of using these models is that
very large mass ratios are predicted for the long period
superhumpers. At such large mass ratios the resonance lies well beyond
the truncation radius of the disc and eccentricity cannot be excited.

We show that when the correct expression for dynamical precession is
used, there is poor agreement with the observations. This can be seen
even though there is considerable uncertainty in estimating the mass
ratio of a system of given orbital period.

The observed superhump periods of the eclipsing systems OY~Car, Z~Cha
and HT~Cas are shown to be significantly less than predicted by the dynamical
theory, demonstrating that the retrograde contribution of pressure
forces is important. 

The inclusion of a retrograde pressure contribution to the precession
rate not only improves the fit to the data but also requires long
period superhumpers have smaller mass ratios than previously
thought. This in turn is more consistent with the eccentricity being
generated at the $3:1$ Lindblad resonance. 
 
\section*{Acknowledgements}
The author would like to thank Brian Warner for many useful
discussions; and Steve Lubow and Jim Pringle for helpful remarks. JRM
is employed on a grant funded by the Leverhulme Trust.


\begin{thebibliography}{}
\bibitem{BinneyTremaine}
Binney, Tremaine, 1987, Galactic Dynamics, Princeton University Press,
Princeton, New Jersey
\bibitem{Brouwer}
Brouwer, D., Clemence, G.D., 1961, Methods of Celestial Mechanics, Academic
Press, London
\bibitem{Flebbe}
Flebbe, O., Muenzel, S., Herold, H.; Riffert, H., Ruder, H., 1994,
ApJ, 431, 754
\bibitem{HiroseOsaki}
Hirose, M., Osaki, Y., 1990, PASJ, 42, 135
\bibitem{HiroseOsaki2}
Hirose, M., Osaki, Y., 1993, PASJ, 45, 595
\bibitem{HTCasq}
Horne, K., Wood, J.H., Stiening, R.F., 1991, ApJ, 378, 271
\bibitem{Kunze}
Kunze, S., Speith, R., Riffert, H., 1997, MNRAS, 289, 889
\bibitem{LinPringle}
Lin, D.N.C., Pringle, J.E., 1976, in Eggleton, P., Structure and
Evolution of Close Binary Systems, Reidel, Dordrecht
\bibitem{Lubow91}
Lubow, S.H., 1991, ApJ, 381, 259
\bibitem{Lubow92}
Lubow, S.H., 1992, ApJ, 401, 317
\bibitem{Molnar}
Molnar, L.A., Kobulnicky, H.A., 1992, ApJ, 392, 678
\bibitem{me96}
Murray, J.R., 1996, MNRAS, 279, 402
\bibitem{me98}
Murray, J.R., 1998, MNRAS, 297, 323
\bibitem{meBrianDayal}
Murray, J.R., Warner, B., Wickramasinghe, D.T., 1999, MNRAS, submitted
\bibitem{Darrah}
O'Donoghue, D., Charles, P.A., 1996, MNRAS, 282, 191
\bibitem{Patterson}
Patterson, J., 1998, PASP, 110, 1132
\bibitem{Wood}
Simpson, J.C., Wood, M.A., 1998, ApJ, 506, 360
\bibitem{SmithDhillon}
Smith, D.A., Dhillon, V.S., 1999, MNRAS, in press
\bibitem{Rob1}
Whitehurst, R., 1988, MNRAS, 232, 35
\bibitem{Robonly}
Whitehurst, R., 1994, MNRAS, 266, 35
\bibitem{RobAndrew}
Whitehurst, R., King, A.R., 1991, MNRAS, 249, 25 
\bibitem{OYCarq}
Wood, J.H., Horne, K., Berriman, G., Wade, R.A., 1989, ApJ, 341, 974
\bibitem{ZChaq}
Wood, J.H., Horne, K., Berriman, G., Wade, R.A., O'Donoghue, D.,
Warner, B., 1986, MNRAS, 219, 629

\end{thebibliography}
\end{document}